\documentclass[%
 reprint,
 aip,
 graphicx,
 amsmath,amssymb,
]{revtex4-2}
\usepackage[pdftex]{graphicx}
\usepackage{dcolumn}
\usepackage{bm}

\usepackage{color}
\usepackage[flushleft]{threeparttable}

\usepackage[hidelinks]{hyperref}
\usepackage[capitalise]{cleveref}
\usepackage[utf8]{inputenc}
\usepackage[T1]{fontenc}
\usepackage{mathptmx}
\usepackage{etoolbox}
\makeatletter
\def\@email#1#2{%
 \endgroup
 \patchcmd{\titleblock@produce}
  {\frontmatter@RRAPformat}
  {\frontmatter@RRAPformat{\produce@RRAP{*#1\href{mailto:#2}{#2}}}\frontmatter@RRAPformat}
  {}{}
}%
\makeatother

\usepackage{soul}
\definecolor{lightgray1}{gray}{0.9}

\usepackage[most]{tcolorbox}

\tcbset {
  base/.style={
    arc=0mm, 
    bottomtitle=0.5mm,
    boxrule=0mm,
    colbacktitle=black!10!white, 
    coltitle=black, 
    fonttitle=\bfseries, 
    left=2.5mm,
    leftrule=1mm,
    right=3.5mm,
    title={#1},
    toptitle=0.75mm, 
  }
}

\definecolor{brandblue}{rgb}{0.34, 0.7, 1}
\newtcolorbox{mainbox}[1]{
  colframe=brandblue, 
  base={#1}
}

\newtcolorbox{subbox}[1]{
  colframe=black!30!white,
  base={#1}
}

\begin{document}

\preprint{AIP/123-QED}

\title[Shape deformation of magnetically levitated fluid droplets]{Shape deformation of magnetically levitated fluid droplets}
	\author{I. Sanskriti${}^*$}
	\author{D. Kim}
    \author{J. Twamley }\email{isha.sanskriti1@oist.jp}
\affiliation{ 
Quantum Machines Unit, Okinawa Institute of Science and Technology Graduate University, Onna, Okinawa 904-0495, Japan \looseness=-1
}%

\date{\today}

\begin{abstract} \label{abs}
Diamagnetic levitation can provide a completely passive method to support materials against the pull of gravity, and researchers have levitated both solids and fluids. Such levitation can be assisted by increasing the magnetic susceptibility contrast by using a surrounding paramagnetic medium and through buoyancy forces, known as magneto-Archimedean levitation. The magneto-Archimedean levitation of solids has proved useful in chemistry and biology. However, the levitation of fluid droplets has an additional interest because the fluid droplet's shape can deform. 
We perform experiments and simulations to gauge the squashing or eccentricity of the static magnetically levitated fluid droplet. By carefully characterizing all the parameters affecting the droplet's levitation, using image analysis to estimate the droplet's eccentricity, and using finite element adaptive simulations to find the lowest energy droplet shape, we find good agreement between the simulations and experimental results.  As a potential application, we show that the droplet's eccentricity can be used to perform magnetic gradiometry with a potential resolution of $S\sim 8\,{\rm nT/cm}$, over a volume of 10 mm$^3$, which is competitive with other room-temperature magnetic gradiometer techniques.  


\end{abstract}

\maketitle



Studying the shape of liquid droplets has a long history. \cite{Rayleigh1879VI.Jets, Lamb1892Hydrodynamics} 
Developing a deep understanding of the shape of liquid droplets and their dynamics can reveal many properties, such as the influence of droplet shapes and dynamics in radar sensing of rainfall activity.
 \cite{Hill2012ShapeDroplet} 
The study of droplet shape is not  limited to fluid mechanics but also informs   the study of processes over a wide range of length scales, from the shapes of astrophysical bodies to the fission of atomic nuclei.\cite{Holgate2018ShapesVacuum, Butler2020EquilibriumTension}
 By levitating droplets, one can study their shape over a long duration, and initial studies focused on the acoustic levitation of droplets. \cite{Lee1991StaticDropb, Yarin1998OnDroplets} Researchers even demonstrated the acoustic levitation of liquid metal droplets. \cite{Xie2002LevitationUltrasound} 

Diamagnetic levitation, which does not require active power, is highly suitable for passive static levitation scenarios and has a long development history. \cite{Thomson1850RemarksSubstances, Braunbek1939FreiesMagnetfeld,Berry1997,Simon2000} Researchers demonstrated the diamagnetic levitation of various organic fluids \cite{Beangnon1991LevitationMaterials}, but this required a bitter solenoid or superconducting magnet with the approximate magnetic field of 15-20 T. Researchers could also magnetically levitate liquid helium droplets, but it required $|B(dB/ dz)|$ of $20.7$ T$^2/$cm. \cite{Weilert1996MagneticHelium} Later, by combining buoyancy forces and a paramagnetic medium, researchers levitated fluids at even lower magnetic field strengths of 10 T and named the method magneto-Archimedes levitation.\cite{Ikezoe1998MakingLevitate} These studies opened up a new dimension of diamagnetic levitation and its potential applications in various fields, including
 chemistry, biology, materials science, and biochemistry. \cite{Ge2020MagneticBiochemistry}
The development of magneto-Archimedean levitation of solids and liquids led researchers to study the dynamics, shape, and size of levitated static drops or levitated spinning drops affected by magnetic forces. \cite{Whitaker1998ShapeDrops, Katsuki2007MagneticDrop, Hill2008NonaxisymmetricDroplet,Hill2010VibrationsDroplet, Temperton2014MechanicalDroplets, Liao2017ShapesDrops}
 Researchers also explored this technique to separate biological materials, \cite{Hirota2004Magneto-ArchimedesMaterials}  measure densities of solids and water immiscible organic liquids, \cite{Mirica2009MeasuringFundamentals,Ge2018AxialManipulation} develop density-based sensors to characterize food and water, \cite{Mirica2010MagneticWaterb} and separate and identify illicit drugs. \cite{Abrahamsson2020AnalysisLevitation}

Many of these works focused on studying the shape dynamics of the fluid droplet, while others focused on the levitation height of the fluid droplet. In this work, we focus on determining the deformation in the shape of the magneto-Archimedean levitated droplet due to the combination of magnetic, buoyancy, gravity, and surface tension forces. There have been only a few previous studies on the magnetic shape deformation of diamagnetically fluid drops, but these either consider pendant drops, \cite{Katsuki2007MagneticDrop,Poulose2022DeformationField} or air bubble magnetically levitating within liquid oxygen in a large superconducting electromagnet. \cite{Duplat2013OnEnvironment}

 In this work, we explore if the magnetic forces in a magneto-Archimedean fluid levitation setup can be engineered to affect a significant deformation of the static shape of the levitated droplet. Instead of superconducting magnets, we have used permanent rare-earth magnets and a diamagnetic organic fluid droplet levitated in a paramagnetic aqueous salt medium. Here, we demonstrate that the levitated droplet exhibits significant squashing in its shape and becomes oblate by adjusting the position of the trapping magnets due to the increased strength of the magnetic forces on the droplet in an anisotropic fashion. We performed experiments to measure the shape deformations of the droplet as we altered the magnetic forces, and these agreed with numerical simulation.
 
 Our findings indicated that the droplet eccentricity depends on the magnetic field gradients at the droplet's location, which led our investigation to whether the eccentricity can be used as a sensor for magnetic gradients, e.g., work as a magnetic gradiometer. 
There exist already various types of magnetic field sensors or gradiometers, each with its strengths and weaknesses. 
Through simulations, we found that the estimation of the droplet eccentricity can yield a magnetic gradiometer whose sensitivity is competitive with many existing magnetic gradiometer technologies, operates at room temperature, and is compact and robust.

In the following, we describe the physics of the magneto-Archimedean droplet trap, outline the experimental setup, and then work to carefully quantify each of the parameters and variables that define the trapping forces. 
As a check, we examined theoretically and experimentally the equilibrium droplet levitation heights as we varied the magnetic forces (by moving the magnets), which agreed well with previously published works. \cite{Mirica2009MeasuringFundamentals,Ge2018AxialManipulation} We performed experimental measurements and numerical simulations of the eccentricity of the static droplet's shape. Finally, we examined the potential for the shape deformation to act as a magnetic gradient sensor.

 For describing the physics of trapping, we considered an infinitesimal element of the diamagnetic fluid of volume $V$, located at position $\vec{r}$. We can express the total net force density it experiences as \cite{Mirica2009MeasuringFundamentals},  $\vec{F}(\vec{r})/V\equiv \vec{f}=-\vec{\nabla} u(\vec{r})$, and the potential density $u(\vec{r})$ is given as 
\begin{equation}
    u(\vec{r})=(\rho_s-\rho_m)g\,z -\frac{1}{2\mu_0}(\chi_s-\chi_m)|\vec{B}(\vec{r})|^2\;\;,
    \label{Eq:upervolume}
\end{equation}
where $(\rho_s, \chi_s)$, is the (density [${\rm kg/m^3}$], volume magnetic susceptibility [unit-less]), of the sample (also called the droplet fluid), while  $(\rho_m, \chi_m)$ is that of the surrounding medium (or paramagnetic fluid). The diamagnetic sample fluid, 3-chlorotoluene, has known values of $\rho_s$ and $\chi_s$ and was used as received from the suppliers. For the paramagnetic medium, $\rm{ MnCl_2\cdot 4H_2O}$, the density and magnetic susceptibility will vary depending on the prepared molar concentration of dissolved paramagnetic salts in an aqueous solution. The magnetic field $\vec{B}(\vec{r})$ [T] is generated by two opposing ring magnets whose inter-magnet separation is $d$ [m], and $z$ [m] is the $z-$component of $\vec{r}$, the height of the infinitesimal volume element above the top surface of the bottom magnet. We take the origin of the axes as the axial center of the top surface of the bottom ring magnet.

\begin{figure}[ht!]
\begin{center}	\includegraphics[width=.99\linewidth, bb=0 0 623 264]{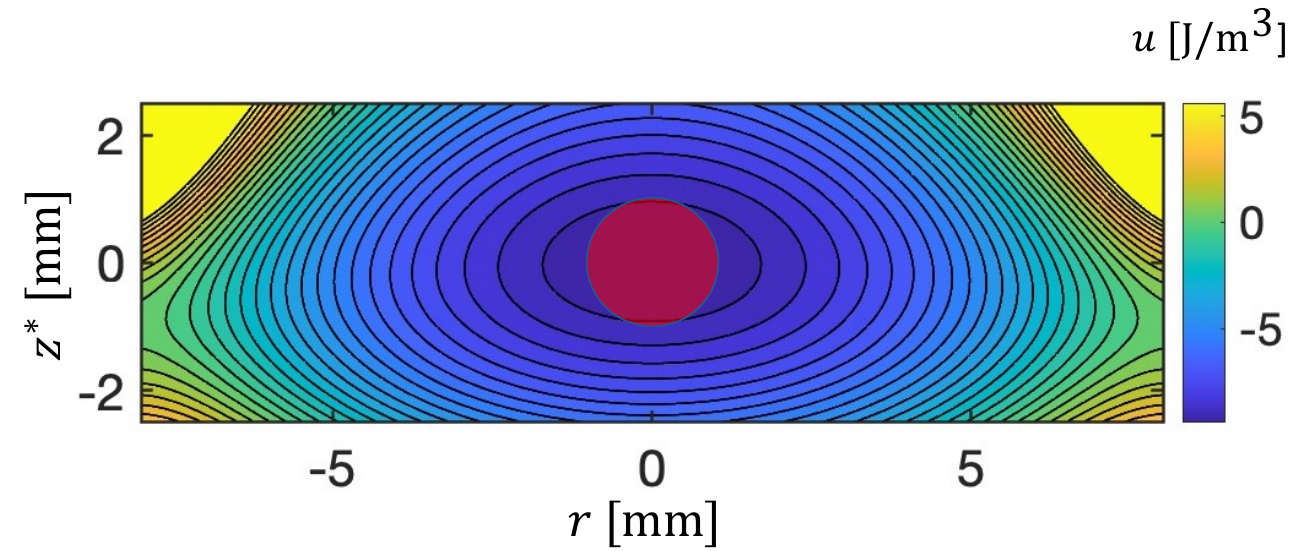}
\end{center}\vspace*{-10pt}
\caption{Example potential density plot. We graph the Archimedean Magnetic potential density given in \cref{Eq:upervolume}, in axial coordinates, which includes the effects of the magnetic, gravitational, and buoyancy forces. We observe a minimum that traps the diamagnetic droplet in space. Here the magnet separation is set as $d=7\,{\rm mm}$, with the droplet volume $V_{drop}=10\,{\rm \mu L}$, and radius $r\sim 1.33\,{\rm mm}$, with a $3\,{\rm M}$ concentration of the paramagnetic medium.  Solving \cref{Eq2:height} yields a levitation height of $h \sim 5.25\,{\rm mm}$, and we choose $z^*=z-h$ to center the plot at this height. The contours are spaced by $\Delta u=0.5\, {\rm [J/m^3]}$. The red disc indicates the approximate position and dimensions of the diamagnetic droplet. Although the potential exhibits significant oblateness, the deformation of the droplet towards these contours is resisted by the high interfacial tension. In this example, we have chosen: magnet dimensions [OD, ID, H]: [38.6, 19.1, 30.9] ${\rm mm}$, the magnetization of the magnets are assumed to be identical $M\sim 1.32\,{\rm T}$. The droplet fluid is considered to be pure 3-chlorotoluene, and the medium fluid is a $3\,{\rm M}$ concentration of aqueous $\rm MnCl_2\cdot 4H_2O$, with the associated physical properties as listed in [Table S1, SI].
}
\label{fig:potential_field}
\end{figure}

We can estimate the equilibrium height, $h$, of the levitated droplet by solving 
\begin{equation}
    f_z(\vec{r}=(0,0,z=h))=0,\;\;\Rightarrow\;\;\left.\frac{d}{d z}\,u(\vec{r}=\{0,0,z\})\right|_{z=h}=0\;\;.\label{Eq2:height}
\end{equation}
This approximation is justified when the droplet volume is infinitesimal so that $\int_V f_z(\vec{r}) \approx 0$.
In our case, the droplet size is comparable to the inter-magnetic separation $d$, and we find that the magnetic potential around the droplet is non-Gaussian. Therefore, we can use the following to obtain a more precise estimation of the equilibrium droplet height, $h$, by solving 
\begin{equation}
    F_z(\vec{r}=(0,0,z=h))=0,\;\;\Rightarrow\;\;\left.\frac{d}{d z}\left( \,\int_{V(z,r)} u(\vec{r})\,dV\right) \right|_{z=h}=0\;\;.\label{Eq2:height_vol_int}
\end{equation}
where $V(z,r)$ denotes a spherical integration volume of radius $r$, the droplet radius, where the sphere is centered at the position $\vec{r}=\{0,0,z\}$. As an example, we show the potential density $u(r,z^*)$, in \cref{fig:potential_field}, where we set $z^*=z-h$.

To evaluate the equilibrium shape of the droplet when it is levitating in the potential density $u(\vec{r})$, we also require the value of the interfacial tension (IFT), $\gamma$ [N/m], between the droplet fluid and the medium fluid. 
We see that the small changes in $\gamma \sim \pm 2$ mN/m result in changes in the droplets' eccentricity by a few parts in a hundred. However, if the change in $\gamma \sim \pm 30$ mN/m, the change in eccentricity can be a few parts in ten.

\begin{figure*}[tb]
\centering
\setlength{\unitlength}{1cm}
\begin{picture}(8,7)
\put(-4,-.2){\includegraphics[width=.9\linewidth]{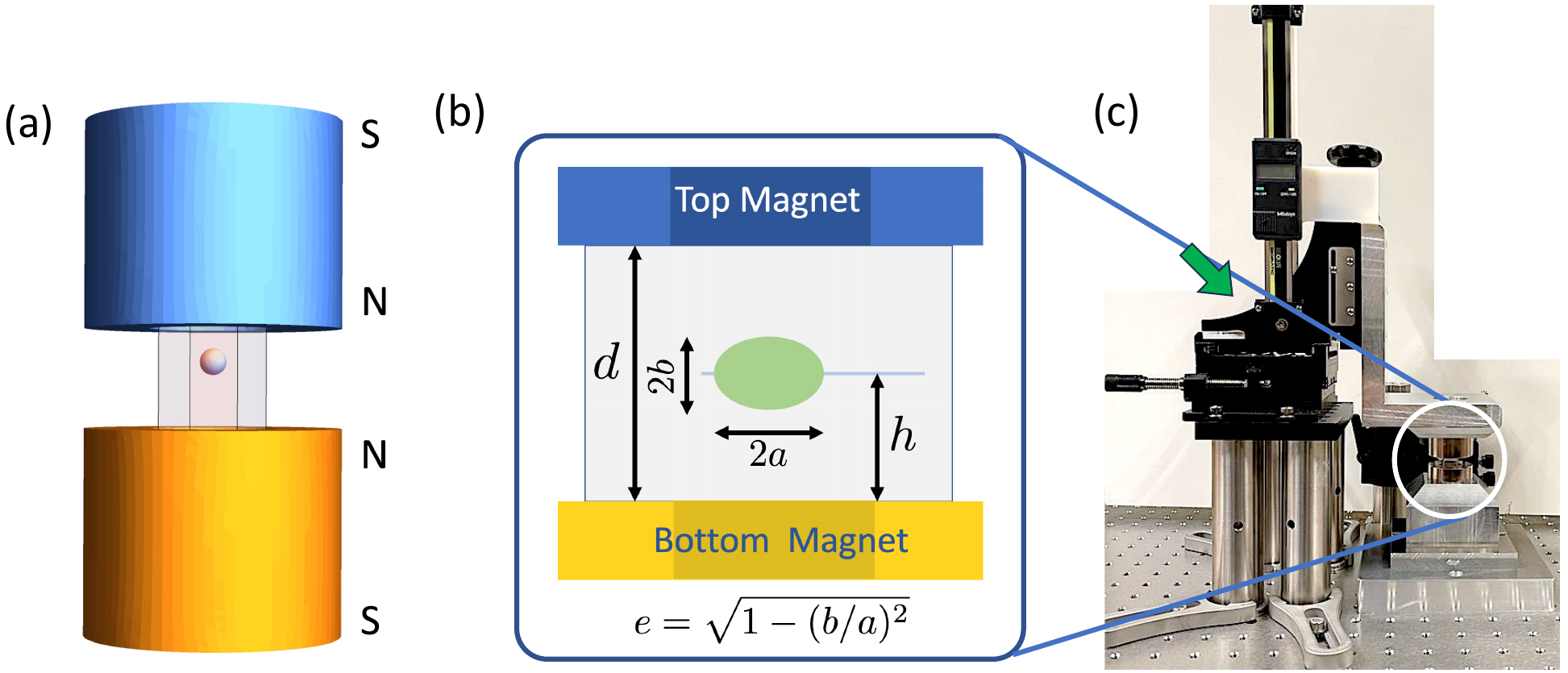}}
\end{picture}
\caption{Schematic of magnetic levitation and squashing of droplet: (a) a small (1-2 mm diameter) droplet of a diamagnetic liquid (round sphere) is trapped by a combination of magnetic and buoyancy forces within a rectangular glass cuvette, containing a paramagnetic liquid (clear). The cuvette sits between two opposing NdFeB ring magnets, producing an anti-Helmholtz field. The separation distance, $d$, between these magnets can be varied. (b) detail of the squashed levitated droplet (green - exaggerated for clarity), with a levitation height $h$, magnet separation $d$, and major (minor) droplet radii $a\,(b)$, and eccentricity $e$; (c) photo of the experimental setup illustrating the ring magnets and reinforced micrometer  positioning arrangement of the upper ring magnet indicated by a green arrow, and micrometer gauge to measure the magnet separation. Not shown are the rigs to hold the microscope camera to measure the droplet shape, lights to illuminate the droplet, and micrometer Hall-probe scanning stage to measure the trapping $B$ fields.}

\label{Fig1:Schematic}
\end{figure*}

 We consider the setup depicted in  \cref{Fig1:Schematic}, which is similar to that in \cite{Ge2018AxialManipulation}, consisting of two opposed NdFeB (N40 - Neomag) ring magnets with dimensions $38.6\,{\rm mm}$ (outer diameter) $\times\,19.1\,{\rm mm}$ (inner diameter) $\times\, 30.9\,{\rm mm}$ (height), surrounding a rectangular quartz cuvette containing a paramagnetic solution of $\rm MnCl_2\cdot 4H_2O$, prepared with various concentrations. A fixed-volume droplet of diamagnetic liquid (3-chlorotoluene) was inserted into the paramagnetic medium and was trapped by a combination of buoyancy and magnetic forces. As indicated by the green arrow in \cref{Fig1:Schematic}(c), we can adjust the position of the upper magnet and, in particular, the inter-magnet separation $d$ via a robust micrometer stage.  This stage has to withstand considerable repulsive forces generated between the magnets when $d$ is reduced. The setup shown also has affixed a digital micrometer to record the inter-magnet separations accurately. 
 For measuring the levitation height and capturing the image of the levitated droplet between the two magnets, a microscope (Dino-lite Edge 3.0), along with lights, was mounted to illuminate the levitated droplets.


To precisely utilize \cref{Eq:upervolume}, we must have precise information about the densities and magnetic susceptibilities of the sample and medium and the magnetic fields. The preparation of various concentrations of the paramagnetic medium solution is described in [Sec III, SI]. 
The density of the medium solution $\rho_m(T,c)$ is a function of temperature and concentration of the paramagnetic salt $c$. We measured $\rho_m$ for various values of $c$ using a pycnometer [Sec IV, SI]. The experimental values of density of (1 M, 2 M, and 3 M) aqueous solutions of MnCl$_2\cdot$4H$_2$O at 25 $^\circ$C were measured to be (1.097 $\pm$ 0.004, 1.195 $\pm$ 0.005, 1.293 $\pm$ 0.005) g/mL respectively that compared well with the empirical formula [Eq S4, SI].
The magnetic susceptibility of the {{3-chlorotoluene}} $\chi_s$ is taken from the literature, while $\chi_m(T,c)$ for the paramagnetic medium is a function of temperature and salt concentration. Values for $\chi_m(T,c)$ are calculated using the Curie-Weiss law [Sec V, SI].
The potential density is also a function of the magnetic fields generated by the two ring magnets. By setting the ring magnets to have a separation of $d=12$ mm, we experimentally measured the $B_z(z)$ field along the $z-$axis between the two ring magnets using a Hall probe magnetometer. By fitting the experimental data to theory, we estimated the magnetization of the top and bottom ring magnets ($M_{top}$ = 1.36 T and $M_{bottom}$ = 1.3 T), which are quite comparable to those quoted by the manufacturer [Sec VI, SI]. These values were used for all subsequent simulations and modeling.
To check the accuracy of the parameters and setup, we measured the levitation height of the droplet for various inter-magnet separations $d\sim 6-12$ mm and compared the experimental data with simulations and found quite good agreement [Sec VII, Fig S5, SI]. The slight systematic upwards shift of the experimental data compared to the theoretical predictions may be due to a slight tilting of the movable ring magnet away from an exactly horizontal orientation.

The droplet's shape will be greatly affected by the value of the interfacial tension (IFT) $\gamma$ between the droplet fluid and the surrounding medium. Hence, we determined the IFT between 3-chlorotoluene and a 3 M aqueous solution of MnCl$_2\cdot$4H$_2$O, $\gamma_{CT/MnCl_2}$, experimentally with the help of Pendant drop Tensiometry. \cite{Berry2015MeasurementTensiometry} The value for $\gamma_{CT/MnCl_2}$ was found to be $45.23\pm 0.55\, {\rm  mN/m}$. This value is used later in simulations of the droplet shape. The experimental details, along with the figure of the pendant droplet, are given in [Sec VIII, SI].

\begin{figure}[ht!]
\centering
\includegraphics[width=1\linewidth]{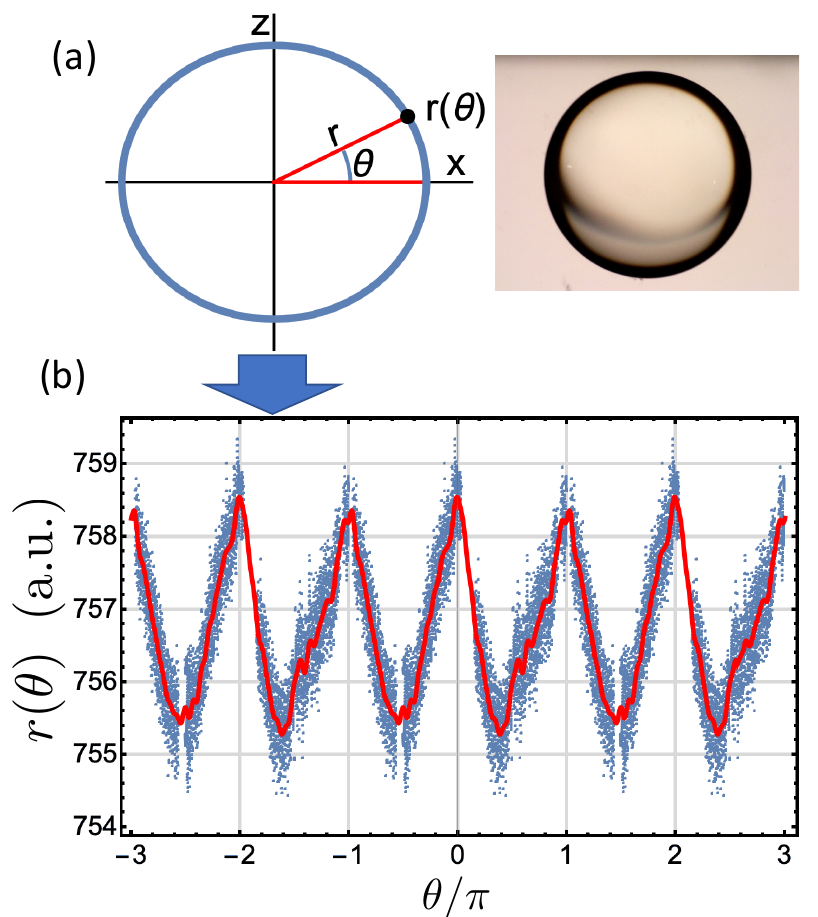}
\vspace*{-10pt}\caption{
Image analysis of a droplet shape for inter-magnet separation $d=9$ mm. (a) [Left] we schematically depict how the radius of the droplet depends on the angle $\theta$, (a) [Right] actual microscope image of the levitated drop, (b) the analysis of the droplet image in (a) [Right], where the blue points label points on the edge of the droplet and the data is repeated outside the fundamental domain $\theta\in [0,2\pi]$. The red curve is a fitted function for $r(\theta)$, using a high-order Fourier series. From the minimum and maximum of $r(\theta)$ and the spread of values around these, we estimate the eccentricity of this droplet to be $e=0.093\pm 0.018$.
 }
\label{fig:polardata}
\end{figure}

We performed experiments and image analysis using Mathematica to determine the eccentricity of the droplet with varying inter-magnet separation $d$.
A digital microscope was used to take close-up photos of the levitated droplet. 
The lighting conditions were adjusted to capture photos where the droplet edge exhibited high contrast with its background. 
Two side light sources illuminated the levitated droplet. To achieve high droplet edge contrast in the images, we placed a black square piece of hardboard with a circular central white spot (2 cm in diameter), which is placed in the background behind the droplet as seen by the microscope. This leads to an image where the background and the droplet interior are white, but the droplet edge is dark. This results in a high-contrast image of the edge of the droplet. 
As the inter-magnet separation distance was changed, several microscope images of the droplet were taken.  
Image processing was performed on the high-contrast edge of the droplet to identify the coordinates of points. From this, an estimation of the eccentricity of the droplet image was obtained. The droplet image was processed by first binarizing and blackening the interior. This created a higher contrast for further processing for edge detection. Then, the 2D coordinates of points on the edge were isolated. The center point of the droplet was also identified during image processing.  Next, the radial distance from this center point to each point $i$ on edge and angle to the horizontal was found, e.g., $(r_i, \theta_i)$, where $\theta_i\in [-\pi,+\pi]$.
The experimental data was found to be periodic in $\theta$ with period $2\pi$, and we used this periodicity to expand the domain of $\theta\in [-3\pi,+3\pi]$. We dropped some data points clustered around $\theta\sim m\pi/2$, where $m$ is an integer, as the finite  resolution of the camera image, which is in a cartesian grid, does not map very smoothly under the polar transformation at these angles \cite{Poulose2022DeformationField}.
 We fit a Fourier expansion to this periodic data as $r(\theta)=\sum_{k=0}^{30}\,[A_k \sin(k\pi \theta+\phi)+B_k\cos(k\pi \theta+\phi)]$, fitting for the amplitudes $A_k,\,B_k$, and phase shift $\phi$. The latter parameter describes any slight mismatch of the microscope photo's vertical axis from the true lab vertical axis.  From this fitting, we found the major axis $a=\max( r(\theta))$ and minor axis $b=\min(r(\theta))$ of the droplet. We estimated the error in these quantities, $(\delta a, \delta b)$, by the standard deviation of $r_i$ around these locations. An example plot of $r_i$ vs. $\theta_i$, for a drop with $d=9$ mm, is shown in \cref{fig:polardata}. From this, we observe that these errors are not small compared to the mean value of the eccentricity. One can find the error in the derived eccentricity $e=\sqrt{1-(b/a)^2}$, as
\begin{equation}
    \delta e=\left(\frac{b^2}{a^3}\right)\frac{\delta a}{e}+
    \left(\frac{b}{a^2}\right)\frac{\delta b}{e}\;\;.
\label{eq:errorineccentricity}
\end{equation}
As the inter-magnet separation distance $d$, was varied from $d=7, 8, 9, 10, 11$ mm, for each separation, we took several photo-micro-graphs of the droplet, performed the above-described image analysis, and for each image, obtained an estimate for the eccentricity, and its error $e\pm \delta e$.
In the cases when $d>11$ mm and $d<7$ mm, the image analysis failed as the droplet was partially obscured by the magnets [Sec IX, SI].





In the subsequent sections, we describe how we model and predict the shape of the levitated droplet numerically, in particular, estimate the droplet's eccentricity. 
Since the magnetic fields produced by the ring magnets are axially symmetric, the droplet shape will also be axially symmetric (i.e., the shape will be independent of $\phi$), and we can thus work in a reduced two-dimensional $x-z$ plane.
We numerically sampled $u(\vec{r})$ in a 2D region surrounding the droplet on this plane using \cref{Eq:upervolume}, where the $B$ field produced by the magnets is simulated using MagPyLib. \cite{Ortner2020Magpylib:Computation}
The shape of the droplet is that which minimizes the surface energy in the presence of $u(\vec{r})$. Two limiting cases are when a) if the interfacial tension vanishes $\gamma = 0$, the shape of the droplet exactly follows the equipotentials of the potential density $u(\vec{r})$, or b)  if $\gamma\sim \infty$, then the shape of the droplet is a sphere.
However, in the case of finite $\gamma$, estimating the droplet's shape is not straightforward. Such surface energy minimization calculations can be performed using the Surface Evolver (SE). \cite{Brakke1992TheEvolver, Brakke1997UsingFoams, Zhang2016StabilityConfiguration} SE can estimate droplet shape through iteration and refinement, including additional force effects and a spatially varying potential energy density. It also considers the interfacial tension and the volume of the droplet enveloped by the surface.

\begin{figure}[h!]
\begin{center}	\includegraphics[width=.9\linewidth]{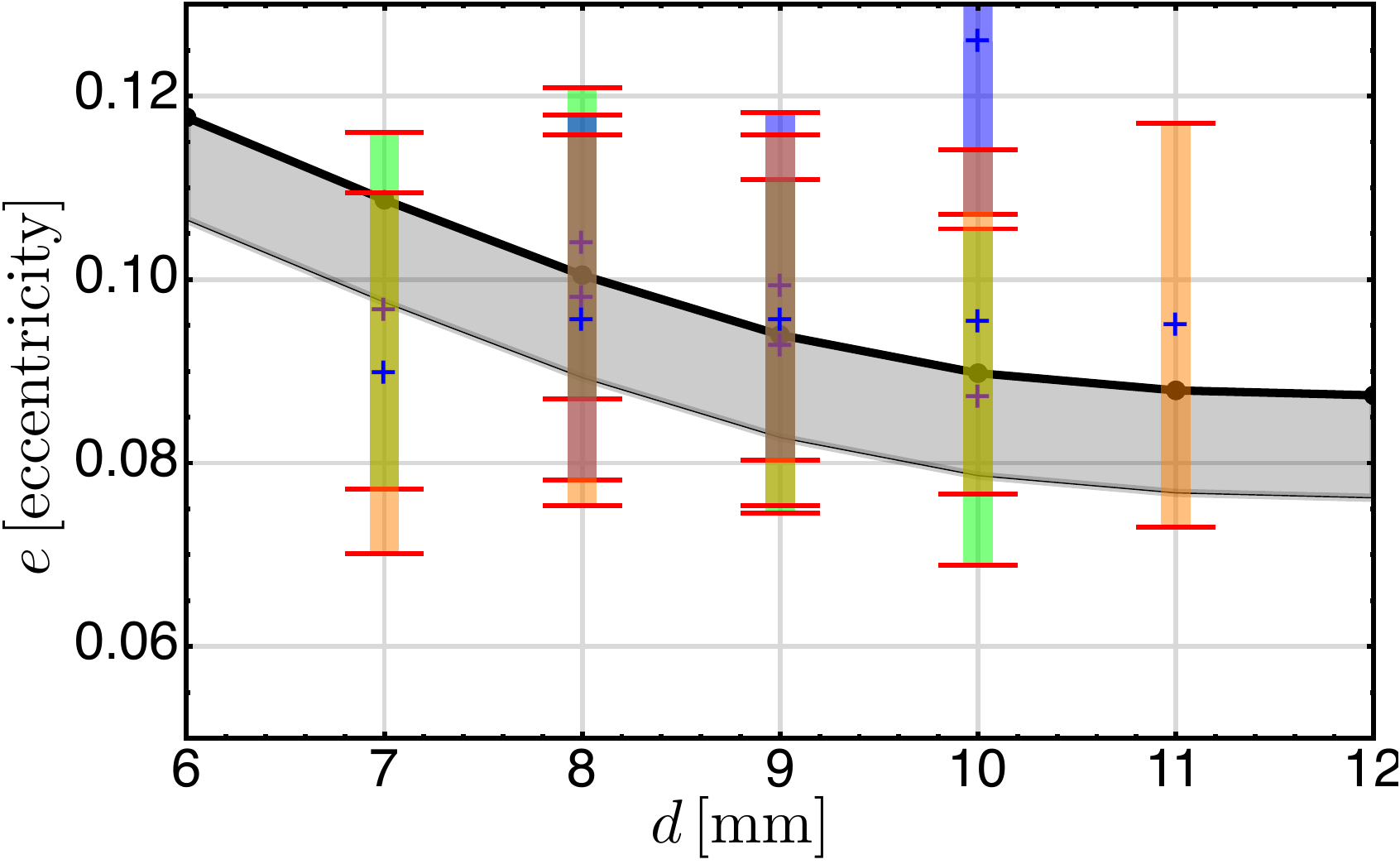}
\end{center}\vspace*{-10pt}\caption{Comparison between simulation and experimental data of eccentricity.
Parameters used for the simulation: $\gamma$ = 45 mN/m, $V$=10 $\mu$L, $c$ = 
3 M. We plot the mean eccentricity (plus markers) and errors for the experimentally determined droplet eccentricity for inter-magnet separations $d=7,8,9,10,11$ mm. The numerical simulation is the solid black line which is an upper bound for the estimated numerical eccentricity. The associated numerical error with this computation is shaded in gray. 
The simulations and experimental data agree except for the purple outlier at $d=10$ mm, which suffered from poor background non-uniformity.
}
\label{fig:eccentricity}
\end{figure}

The Surface Evolver program incorporates spatial potentials as analytic formulae. Thus we sampled the potential density $u(\vec{r})$ around the small droplet and fitted as a polynomial function of degree 2 in the coordinates $x$ and $z$, where the origin is taken as the droplet's equilibrium levitation height.
We then called the SE program using this potential, incorporating the interfacial surface tension and droplet volume used in all experiments. 
After appropriate iterations and refinements, vertices of the evolved shape are exported for further analysis. 
%
%
%
%
From the SE exported data, we calculated the radii of all surface points to the geometric center of the 3D surface and set the major and minor radii of the oblate droplet to be the maximum and minimum radii found. From this, we computed the droplet eccentricity. As a study, we performed simulations for varying magnet separation distance $d\in \{4,16\}\,{\rm mm}$ and for various test values of interfacial tensions, and these results are reported in [Fig S10, SI]. As one might expect, the droplet's eccentricity is increased with lower interfacial tensions $\gamma$ and lower magnet separation distance $d$. We also observed that the simulated eccentricity value tends to be a constant for large separations $d$ greater than 10 mm. In this case, the magnetic forces from a single-ring magnet are sufficient to trap the droplet. Since in magneto-statics, one must have $\vec{\nabla}\cdot \vec{B}(\vec{r})=0,\,\forall\,{\vec{r}}$,  these trapping forces cannot be isotropic, and there is some residual squashing of the droplet.
We simulated the droplet's shape in the actual experiments we performed. We plotted $e(d)$, for $V=$ 10 ${\rm \mu L}$, using the experimentally determined value of the interfacial tension $\gamma=45\,{\rm mN/m}$,  as the black curve in \cref{fig:eccentricity}. We also developed extensive methods to estimate the accuracy of the numerical simulations as SE's algorithm is somewhat stochastic [Sec X, SI]. We found that SE returns an estimate for the eccentricity, which is always lower than the actual value. For the experiment, we depict the errors in the numerical simulation in \cref{fig:eccentricity} as a grey region {\em below} the black solid line.

In \cref{fig:eccentricity}, we plot both the experimental and numerical simulation results for the droplet eccentricities, and this summarizes the main result of our inquiry. 
We note that all the droplets are oblate, even for inter-magnet separations where the droplet is trapped by only one magnet. 
We observed that the numerical simulations, including error, agree with the experimental measurements predicting a slight increase in eccentricity as the inter-magnet separation is decreased below 9 mm, which is apparent in the experimental data for $d>7$ mm [See Sec XI, SI for details]. 

We also investigated the potential of utilizing eccentricity as a {\em sensor} of the local magnetic field gradient. 
Different magnetic gradiometry methods have varying sensitivities and applications due to their characteristics.
The Hall sensor has relatively low sensitivity, but due to its versatility, it is commonly used in daily life, such as in smartphones.
On the other hand, sensors using microelectromechanical systems (MEMS), superconductors, and atoms have high sensitivities, but their usage is limited due to the technical specifics required by these techniques. 
From \cref{Eq:upervolume}, we observe that the magnetic squashing force, and thus the eccentricity of the droplet, depends on the gradient of the local magnetic fields. We ask whether the eccentricity can be used as a {\em sensor} of the magnetic field gradient $S$ [T/m], e.g. if we measure the major and minor radii of the droplet to a preset precision, e.g., $\delta r~\sim 1$ pm, what is the minimum resolvable change in the magnetic field gradient that can be sensed by estimating the eccentricity? 
To estimate $S$, we make use of the SE and numerically estimate 
\begin{equation}
    S(d_0) = \left.\frac{\text{d}B_z^\prime(d)}{\text{d}a(d)}\right|_{d=d_0}\times a_{min}\;\;.
    \label{Eq12:sensitivity}
\end{equation}
where we assume the resolution to measure $a$ is $a_{min}\sim 10^{-12}$ m, which is possible using laser interferometry (for example, the Picoscale interferometer).
Using an interfacial tension value of 1 mN/m simulations indicated one can achieve $S\sim$ 8 nT/cm [Sec XII, SI].  The predicted values of $S$ are comparable with other sensitive magnetic gradiometer technologies, which can operate in ambient conditions and have a volume of 10 mm$^3$. 

The authors acknowledge support by the Okinawa Institute of Science and Technology Graduate University. The authors also acknowledge technical assistance from P. Kennedy from
the OIST Engineering Section. 

The data and simulation codes that support the findings of this study are available from the corresponding author upon reasonable request.

\bibliography{references_liquid_levitation}

\end{document}